\documentstyle[prd,aps,epsfig]{revtex}
\begin{document}
\thispagestyle{empty}

\begin{center}
\parbox{17.5cm}
{
\begin{flushright}
LPT-ORSAY 00/38\\ 
HD-THEP-00-11\\
PCCF RI 0007\\
\end{flushright}
\vspace*{1.5cm}

\begin{center}
{\large\bf Semileptonic inclusive heavy meson decay:\\[0.2cm]
duality in a nonrelativistic potential model in the Shifman-Voloshin limit}

\vspace*{1.5cm}

{\sc A. Le Yaouanc}$^{a}$, {\sc D. Melikhov}$^{b*}$, 
{\sc V. Mor\'enas}$^{c}$, {\sc L. Oliver}$^{a}$, 
{\sc O. P\`ene}$^{a}$, {\sc J.-C. Raynal}$^{a}$
\end{center}

\small{
\vspace*{1cm}
\hspace*{.8cm}
${}^a$ {\it Laboratoire de Physique Th\'eorique, Universit\'e de Paris XI, 
B\^atiment 210, 91405 Orsay Cedex, France${}^{**}$}

\vspace*{.1cm}
\hspace*{.8cm}
$^b$ {\it Institut f\"ur Theoretische Physik, Universit\"at Heidelberg,
Philosophenweg 16, D-69120, Heidelberg, Germany}

\vspace*{.1cm}
\hspace*{.8cm}
$^c$ {\it Laboratoire de Physique Corpusculaire, Universit\'e Blaise Pascal - 
CNRS/IN2P3, 63000, Aubi\`ere Cedex, France} 
}
}
\end{center}

\vspace*{1.5cm}

\begin{center}
\parbox{15cm}{

\centerline{\bf Abstract}
\vspace*{.3cm}
Quark-hadron duality in the inclusive semileptonic decay $B\to X_c l\nu$  
in the Shifman-Voloshin limit $\Lambda \ll \delta m=m_b - m_c \ll m_b, m_c$ is studied 
within a nonrelativistic potential model. The integrated semileptonic decay 
rate is calculated in two ways: first, by constructing the 
Operator Product Expansion, and second by a direct summation of the exclusive 
channels. Sum rules (Bjorken, Voloshin, etc.) for the potential model are 
derived, providing a possibility to compare the two representations for 
$\Gamma(B\to X_c l\nu)$. An explicit difference between them referred to as 
duality-violation effect is found. The origin of this effect is related to 
higher charm resonances which are kinematically forbidden in the decay process 
but are nevertheless picked up by the OPE. 

\vspace*{.2cm}
Within the considered $1/m_c^2$ order the OPE and the sum over exclusive 
channels match each other, up to the contributions of higher resonances, 
by virtue of the sum rules. In particular this is true for the terms of order 
$\delta m^2/m_c^2$ and $\Lambda \delta m/m_c^2$ which are present in each of the 
decay channels and cancel in the sum of these channels due to the Bjorken and 
Voloshin sum rules, respectively. The size of the duality violation effects is 
estimated to be of the order $O(\Lambda^{2+b}/m_c^2\delta m^b)$ with $b>0$  
depending on the details of the potential. 
Constraints for a better accuracy are discussed.

\vskip 15pt PACS numbers: 13.20.He, 12.39.Pn, 12.39.Jh
}
\end{center}

\vspace{3.5cm}
{\small 
$\;{}^{*}$ Alexander-von-Humboldt fellow. On leave from Nuclear Physics Institute,
Moscow State University, Moscow, Russia 

${}^{**}$ Unit\'e Mixte de Recherche - CNRS - UMR 8627}

\newpage
\thispagestyle{empty}
\parbox{17cm}
{\mbox{}}
\setcounter{page}{0}
\newpage
\title{Semileptonic inclusive heavy meson decay:\\
duality in a nonrelativistic potential model in the Shifman-Voloshin limit}
\author{A. Le Yaouanc$^{a}$, 
D. Melikhov$^{b}$\footnote{Alexander-von-Humboldt fellow. 
On leave from Nuclear Physics Institute, Moscow State University, Moscow, Russia}, 
V. Mor\'enas$^{c}$, L. Oliver$^{a}$, O. P\`ene$^{a}$ and  J.-C. Raynal$^{a}$}
\address{
${}^a$ Laboratoire de Physique Th\'eorique, Universit\'e de Paris XI, 
B\^atiment 210, 91405 Orsay Cedex, France\footnote{
Unit\'e Mixte de Recherche - CNRS - UMR 8627}\\
$^b$ Institut f\"ur Theoretische Physik, Universit\"at Heidelberg,
Philosophenweg 16, D-69120, Heidelberg, Germany\\
$^c$ Laboratoire de Physique Corpusculaire, Universit\'e Blaise Pascal - CNRS/IN2P3, 63000
Aubi\`ere Cedex, France} 
\maketitle

\begin{abstract}
Quark-hadron duality in the inclusive semileptonic decay $B\to X_c l\nu$  
in the Shifman-Voloshin limit $\Lambda \ll \delta m=m_b - m_c \ll m_b, m_c$ is studied 
within a nonrelativistic potential model. The integrated semileptonic decay 
rate is calculated in two ways: first, by constructing the 
Operator Product Expansion, and second by a direct summation of the exclusive 
channels. Sum rules (Bjorken, Voloshin, etc.) for the potential model are 
derived, providing a possibility to compare the two representations for 
$\Gamma(B\to X_c l\nu)$. An explicit difference between them referred to as 
duality-violation effect is found. The origin of this effect is related to 
higher charm resonances which are kinematically forbidden in the decay process 
but are nevertheless picked up by the OPE. 

Within the considered $1/m_c^2$ order the OPE and the sum over exclusive 
channels match each other, up to the contributions of higher resonances, 
by virtue of the sum rules. In particular this is true for the terms of order 
$\delta m^2/m_c^2$ and $\Lambda \delta m/m_c^2$ which are present in each of the 
decay channels and cancel in the sum of these channels due to the Bjorken and 
Voloshin sum rules, respectively. The size of the duality violation effects is 
estimated to be of the order $O(\Lambda^{2+b}/m_c^2\delta m^b)$ with $b>0$  
depending on the details of the potential. 
Constraints for a better accuracy are discussed. 
\end{abstract}
\section{Introduction}
The interest in inclusive decays of heavy mesons is two-fold: 
experimental study of such decays can provide important information on the
weak mixing angles of heavy mesons, and a theoretical
treatment of such processes which includes also nonperturbative effects is possible.  
The theoretical framework based on combining the Operator Product Expansion (OPE) and
Heavy Quark (HQ) expansion  provides decay rates and differential distributions
as series in inverse  powers of the heavy quark mass with the coefficients
proportional to the  matrix elements of the operators of a proper dimension 
\cite{cgg,bsuv93,mw,fls}. A remarkable property of this
expansion is that in the leading-order  this is just the free-quark decay, and
the first correction appears only  at order $1/m_Q^2$. 

On the other hand it is understood that the quark-hadron duality  technically
implemented through the OPE is an approximate framework \cite{shifman}. For
example, the calculation based on OPE does not take into account all details 
of the hadron spectrum which lead to the dependence of the set of open decay 
channels on the momentum transfer. The OPE ignores this fact and this 
inevitably yields some errors in the OPE results \cite{isgur}.  

The theoretical description based on the OPE represents the decay rate as a contour
integral in the complex $q^0$-plane (for details see the next section). The OPE
can be justified only in regions of the complex $q^0$-plane away from the
physical region, whereas in the case of the calculation of the decay rate 
(both differential and integrated) the contour always involves a segment which 
is close to the physical
region \cite{cgg}. This can lead to duality-violating effects, i.e. the
difference between the exact and the OPE based results. \par

However, it is not easy to estimate the errors arising in the OPE, since the exact
hadron  spectrum in QCD is complicated and not exactly known. So, 
testing directly the accuracy of the quark-hadron duality is only possible 
in few exceptional cases. Examples discussed in the
literature are QCD in the Shifman-Voloshin (SV) limit \cite{sv}, and the 't Hooft
model \cite{thooft}. 

In the 't Hooft model (2-dimensional QCD with $N_c\to\infty$)  
the spectrum is reduced to an infinite number of single bound states and 
known precisely so that the direct summation of exclusive channels is possible. 
First numerical analysis of the sum over exclusive channels 
reported the presence of the duality-violating $1/m_Q$ correction for the 
total width \cite{gl}. Later the summation was performed analytically for  
the case of a massless light quark \cite{b-thooft}. 
The result of the OPE calculation agreed with the exact result in this case 
through $1/m_Q^4$ order. 

Duality in QCD in the SV limit \cite{sv} has been studied in \cite{bgm,nous}. This
limit requires $\Lambda_{\rm QCD} \ll \delta m = m_b - m_c \ll m_Q$. A peculiar feature of the
SV limit is that a summation over exclusive channels becomes
possible due to kinematical reasons:  the process occurs near the zero recoil and
thus only few decay channels contribute in the leading $1/m_Q$ order. The expansion of
the relevant transition form factors in this kinematical region is known and 
the sum over exclusive channels can be evaluated. The absence of 
$\Lambda_{\rm QCD}/m_Q$ corrections to the free-quark result in the 
semileptonic (SL) decay rate has 
been demonstrated in \cite{bgm}. 
However, to check the absence of $\Lambda_{\rm QCD}/m_Q$ and $\delta m/m_Q$
corrections within the SV kinematics is not enough to ensure duality in the $1/m_Q$ order 
in the general case, beyond the SV limit. Namely, one should also check that potentially 
large terms of order $O\left(\Lambda_{\rm QCD}\delta m^n/m_Q^{n+1}\right)$ which are 
present in individual decay rates cancel in the sum over exclusive channels. 
The analysis of the $\Lambda_{\rm QCD}\delta m/m_Q^2$ terms in the exclusive sum  
was performed in \cite{nous} for QCD in the $V-A$ case. 
It was found that the duality within this accuracy 
requires a new sum rule. The full comparison to higher orders has not yet
been performed.  

We study the quark-hadron duality in the SV limit within a nonrelativistic 
potential model. The model has several features which make it especially 
suitable for this purpose: the model is self-consistent in the SV limit; 
the spectrum of bound states is relatively simple and can be calculated; 
the exact representations of the transition form factors in terms of the hadron 
wave functions are known. 
These features provide a possibility to calculate the exclusive sum. 
We adopt a technical simplification of a Lorentz scalar
current instead of the $V-A$ current, like it is done in ref. \cite{isgur}.

The main purpose of our analysis is to check whether or not the OPE result calculated 
to some order is equal to the sum over exclusive channels expanded to the same 
order. Both series are double expansions in powers of $\Lambda/m_c$ and $\Lambda/\delta m$. 
They are asymptotic series \cite{b-thooft}, and the question of their convergence is left 
for a later publication \cite{15r}. 

Our main results are as follows: 
\begin{itemize}
\item
We construct the expansion of the $T$-product of the two currents  
in a series of local operators (the OPE) in the potential model 
for a general form of the quark potential. Technically this is done by the expansion 
of the Lippmann-Schwinger equation. We consider the expansion to all orders in 
$\Lambda/\delta m$ but neglect terms $\sim \Lambda^n/m_c^n$ with $n \ge 3$. 
This OPE series provides the expansion of the differential and integrated semileptonic 
decay rates in powers of $\Lambda/m_c$ and $\Lambda/\delta m$. 

Let us point out that the OPE series in the potential model has an important distinction 
from the Wilsonian scheme in the field theory: Namely, in QCD (perturbative) contributions 
of small distances below the scale $1/\mu$ is referred to the Wilson coefficients while 
contributions of large distances above this scale is referred to the matrix elements of 
the local operators. As a result both the $c$-number Wilson coefficients and the matrix 
elements of the local operators acquire the $\mu$-dependence. In the potential model we also 
expand the average of the $T$-product of the two current operators over the $B$ meson in a 
series of local operators, but the resulting $c$-number coefficents as well as the average 
values of the local operators (see Eq.(\ref{ope})) do not have a scale dependence.  

\item
The OPE and the sum over exclusive channels are related to each other 
by sum rules, similar to the Bjorken \cite{13r}, the Voloshin \cite{14r}, and 
the whole tower of higher moments \cite{greg}. We derive these
sum rules. They involve an infinite sum of terms corresponding to all hadronic
excitations, with each term having a well-defined heavy mass expansion. 
The question of the heavy mass expansion of the sum (in other words, of the 
uniform convergence of the series) has not been tackled in this paper. 
If the contribution of higher excitations vanishes rapidly enough the uniform 
convergence is expected.

\item
The OPE provides a heavy mass expansion for the inclusive semileptonic decay rate. 
To compare it with the result of summation over exclusive semileptonic 
decay channels we make use of the sum rules. An explicit difference between 
the two expressions is found, both for the integrated and the differential rates. 
This difference corresponds to the contribution of the resonances 
kinematically forbidden in the decay process which are picked up by the OPE. 
This 'unphysical' contribution is related to the poles in
the complex $q^0$-plane outside the physical region which however contribute 
to the OPE result. The size of this duality-violation cannot be 
estimated in all generality since it depends on the potential and on the 
convergence properties of the sums over resonances. 

\item
For the integrated decay rate the OPE prediction and the sum over exclusive 
channels match, {\it up to the duality-violating contributions of higher resonances}, 
within the $1/m_c^2$ order:  
Terms of order $\delta m^2/m_Q^2$, $\delta m\Lambda/m_Q^2$, 
which are present in any individual decay rate cancel in the sum over all channels 
thanks to the Bjorken \cite{13r} and  Voloshin \cite{14r} sum rules, respectively. 
For terms of order $\Lambda^2/m_Q^2$, $\Lambda^3/m_Q^2\delta m$, etc the agreement 
({\it again up to contributions of higher resonances}) is provided by the higher moment  
sum rules. 
The duality-violation induced by the kinematical truncation of these higher resonances 
in general has the order $O(\Lambda^{2+b}/m_c^2\delta m^b)$ where $b$ depends on the 
details of the potential $V(r)$ both at large and small $r$. 

For the smeared differential distributions near maximal $q^2$ 
the violation of the {\it local duality} is found at the order 
$\Lambda \delta m/m_c^2$. 
\end{itemize}

We make an explicit proof of the present results for the special 
case of the harmonic oscillator potential in Ref. \cite{16r}. 
This is important since some demonstrations given below are rather formal.

In the next Section we present some details of the kinematics  and
discuss the analytical properties of the decay amplitude. In Section 3 the
$1/m_Q$ expansion of the quark propagator is performed and  the OPE series for
the SL decay rate in nonrelativistic quantum mechanics is constructed.   In
Section 4 we consider the HQ expansion of the exclusive form factors in  the
potential model,   and derive the inclusive sum rules (Bjorken, Voloshin, etc.)
which are  crucial for comparing the exclusive sum and the OPE result.  In
Section 5 we provide an analytic expression for the duality-violation
contribution and identify its origin. 
We estimate the accuracy of the OPE both for the integrated rate and the
smeared distribution near zero recoil. A special emphasis is laid on discussing
the role of different inclusive sum  rules in establishing the relationship 
between the OPE and the sum over the exclusive channels. Conclusion summarizes
our results. 

\section{Kinematics and the analytical properties of the decay amplitude}
We consider the inclusive SL decay $B\to X_cl\nu$. 
The rate of this process reads 
\begin {eqnarray}
\Gamma(B\to X_c l \nu)&=\frac{1}{2M_B}&\int \frac{d^4q}{2\pi}\theta(q^0>|\vec q|)L(q)W(q)
\end{eqnarray}
where $L$ is the leptonic tensor, and the hadronic tensor $W$ is defined as 
follows 
\begin{eqnarray}
\label{W}
W=\sum_X\int d^4p_X\theta(p_X^0)\delta(p_X^2-M_X^2)\langle B|J|X(\vec p)\rangle
\langle X(\vec p)|J^+|B \rangle\delta_4(p_B-p_X-q).  
\end{eqnarray}
Here the relativistic normalization of states is implied 
\begin {eqnarray}
\langle p|p'\rangle=2p^0(2\pi)^3\delta (\vec p-\vec p').  
\end{eqnarray}

For the sake of clarity we assume the technical simplification that the leptons
are coupled to hadrons through the {\it scalar current}\footnote{
Recall that for the case of the $V-A$ current and massless leptons, 
the leptonic tensor has the form $L_{\mu\nu}\sim g_{\mu\nu} q^2-q_\mu q_\nu$, and 
for the scalar current $L\sim q^2$. We consider throughout the paper the leptonic tensor 
of the generalized form $L=(q^2)^N$.}. In this case the
leptonic tensor is a scalar function of only one variable, $q^2$,  
and the hadronic tensor $W$ depends on the two invariant variables  
$\nu=P_B\cdot q/M_B$ and $q^2$. 
In the rest frame of the $B$-meson these are 
$q^0$ and $q^2$. At $q^0>0$ and fixed $q^2$ the sum in (\ref{W}) runs over the hadronic states 
with masses $M_X<M_B-\sqrt{q^2}$. The decay rate can be written as follows  
\begin {eqnarray} 
\Gamma(B\to X_c l\nu)
=\frac{1}{2M_B}&\int dq^2 dq^0 |\vec q|\theta(q^0>|\vec q|)L(q^2)W(q^0, q^2), 
\end{eqnarray} 
with $q^2=(q^0)^2-\vec q^{\,2}$. 

Equivalently, we can use $q^0$ and $\vec q^{\,2}$. 
Let us consider the $W(q^0,\vec q^{\,2})$ as an analytical function of $q^0$ at fixed
$\vec q^{\,2}$.  One can write the following relation  
\begin {eqnarray}
\frac{1}{2M_B}W(q^0,\vec q^{\,2})=\frac{1}{2\pi i}{\rm disc}_{q^0} T(q^0,\vec q^{\,2})
\end{eqnarray} 
where  
\begin {eqnarray} 
\label{T1} 
T(q^0,\vec q^{\,2})=\frac{1}{2M_B}\int
dx \exp^{-iqx} \langle B|T(J(x),J^+(0))|B \rangle =\frac{1}{2M_B}
\sum_X\frac{|\langle B|J|X(-\vec q)\rangle|^2}{M_X -E_X(-\vec q)-q^0},      
\end{eqnarray}
$E_X(-\vec q)=\sqrt{\vec q^{\,2}+M_X^2}$ is the energy of the state with the mass
$M_X$ and  the total 3-momentum $-\vec q$.  The sum over $X$ in Eq. (\ref{T1}) for
$T$ runs over {\it all} hadron states with the appropriate  quantum numbers. The
selection of the states kinematically allowed in the decay process is  made by
the proper choice of the integration contour in the complex $q^0$ plane. 
Namely, the decay rate $\Gamma(B\to X_c l\nu)$ can be represented as the contour
integral  in the complex $q^0$ plane over the contour $C(\vec q^{\,2})$ which depends on
the value of $\vec q^{\,2}$  (Fig 1) as follows   
\begin {eqnarray} 
\label{ratecontour}
\Gamma(B\to X_c l\nu)&=&\frac{1}{2\pi i}\int d\vec q^{\,2}  |\vec q|
\int\limits_{C(\vec q^{\,2})}dq^0 
\theta\left(q^0>|\vec q|\right)L\left((q^0)^2-\vec q^{\,2}\right) 
T(q^0,\vec q^{\,2}).   
\end{eqnarray}

\begin{figure}[t] 
\begin{center} \mbox{\epsfig{file=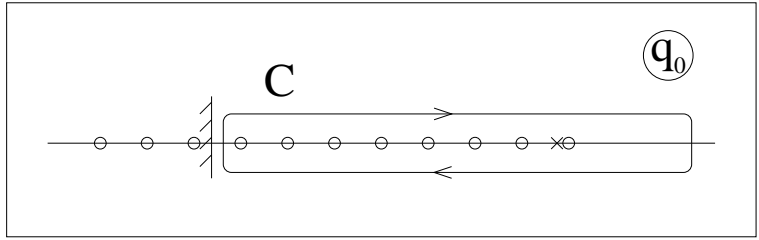,width=10.cm}}
\caption{\label{fig:fig1} Singularities of the amplitude $T(q^0,\vec q^{\,2})$ in
the complex $q^0$ plane. Circles  are hadronic ($\bar c q$) poles which are only
singularities in the  confined potential model, and the cross stands for the
free  $b\to c$ quark process. The vertical line 
${\rm Re}\;(q^0)=|\vec q|$ separates the kinematically allowed region of the real axis 
from the kinematically forbidden region. The contour $C(\vec q^{\,2})$ embraces all
states in the allowed region. Poles at the left of the boundary correspond to
kinematically forbidden bound states.}  
\end{center}
\end{figure}
It is important that the left crossing of the contour $C(\vec q^2)$ with the real axis 
in the complex $q^0$ plane occurs at the point $q_0=|\vec q|$, otherwise the contour 
can be freely deformed in the region where the function $T_0(q_0,\vec q^2)$ is analytic. 
We symbolically mark this constraint with a $\theta$-function in the integrand.  
The integration over such contour selects at any given $\vec q^{\,2}$
only physical states  which can be produced in the decay $B\to X_c l\nu$, i.e.
states   with the invariant masses such that 
$\sqrt{M_X^2+\vec q^{\,2}}<M_B-|\vec q|$. Notice that whereas
the left crossing of the contour with the real axis is tightly fixed at the point 
${\rm Re\,}(q^0)=|\vec q|$, ${\rm Im\,}(q^0)=0$, the right crossing of the contour 
with the real axis can be safely moved to the right. 
In the general relativistic case there are cuts which correspond 
to other physical processes. In the SV limit these
cuts are separated from the physical decay cut by windows of the width $O(m_Q)$. 
In the potential quark model such cuts are absent.

The amplitude $T(q^0,\vec q^{\,2})$ is given by the $T$-product of the two local current
operators,  which is the classical case for performing the OPE. Namely, one has  
\begin{eqnarray} 
\label{ope} 
T(q^0,\vec q^{\,2})= \frac{1}{2M_B i}\int dx
\exp(-iqx)\langle B|T(J(x),J^+(0))|B \rangle=\sum_n C_n(q^0,\vec q^{\,2})\langle B|\hat {\cal O}_n|B \rangle, 
\end{eqnarray}
where $\hat {\cal O}_n$ are local operators and $C_n(q^0,\vec q^{\,2})$ are
the $c$-number coefficients. Introducing the expansion (\ref{ope}) into
(\ref{ratecontour}) gives the integrated rate  as a sum over various local
operators. 

We shall obtain the integrated SL decay rate within our model by the two means: 
first, we construct the OPE series for $T(q^0,q^2)$,  
and second, we calculate directly the sum over exclusive channels.

\section{The model}
We consider this decay in the SV limit,
\begin {eqnarray}
\Lambda\ll\delta m=m_b-m_c\ll m_c,m_b.   
\end{eqnarray}
Notice that in a non-relativistic model, 
$\Lambda$ refers to a fixed energy scale proportional to
the light quark mass $m_d$, to the average quark momentum 
in the hadron rest frame $(\langle B|\vec k^{\,2}|B \rangle)^{1/2}$, to the parameter 
$\beta$ defined in Eq.~(\ref{beta}) and $h_{bd}$ or $\epsilon_B$ to be defined
in Eq.~(\ref{epsb}).
  These parameters may be strongly hierarchized, for example a genuine non-relativistic 
  situation implies 
  $(\langle B| \vec k^{\,2}|B \rangle)^{1/2}\ll m_d$, but all these quantities remain constant
  as $m_c, m_b \to \infty$, they remain proportional
   to some fixed hadronic
  scale which we call $\Lambda$ by analogy with QCD. 
  This is to be distinguished from $\delta m$ which is taken as an independent 
  parameter. Thus we consider the double limit $\delta m/m_c \to 0$,
  and $\Lambda/\delta m\to 0$. Notice finally that $|\vec q|$ is of order
  $\delta m$. 
  
  To avoid confusion, it is important to stress that 
  the standard OPE expansion assumes $\delta m/m_Q$ constant, even if small. 
  So the order of a term $O(\Lambda^{n+m}/m_c^n (\delta m)^m)$ in this paper 
  corresponds to the order $O(\Lambda^{n+m}/m_Q^{n+m})$ of the standard OPE 
  expansion. 
 
We treat the leptonic part relativistically, but for the calculation of the 
hadronic tensor we use the nonrelativistic potential model.  The
nonrelativistic treatment of the hadronic tensor is consistent within the SV
kinematics  and can be used as a tool for studying some of the aspects
of quark-hadron duality. We shall make use the fact that in the
nonrelativistic potential model we know  the structure of the hadron spectrum and
have an exact representation for the hadronic matrix elements of the quark
currents. 

It is convenient to use the nonrelativistic normalization 
of states (which is used hereafter) 
\begin {eqnarray}
\langle p|p'\rangle=(2\pi)^3\delta (\vec p-\vec p'),   
\end{eqnarray}
and consider the process in the rest frame of the decaying $B$-meson.   
The $B$ meson is the ground eigenstate of the Hamiltonian $\widehat H_{bd}$, 
\begin{eqnarray}
\label{eigen}\label{epsb}
\widehat H_{bd}|B \rangle=M_B|B \rangle=(m_b+m_d+\epsilon_B)|B \rangle.
\end{eqnarray}
In the $B$-rest frame this Hamiltonian has the form  
\begin{eqnarray}\label{hbd}
\widehat H_{bd}=m_b+m_d+\frac{\vec k^{\,2}}{2m_b}+\frac{\vec k^{\,2}}
{2m_d}+V_{bd}(r)\equiv m_b+m_d+h_{bd}.    
\end{eqnarray}

For the $B\to X_c$ transition we need the 
$c\bar d$ bound states with the total 3-momentum $-\vec q$, which we denote 
$D_n(-\vec q)$. These are eigenstates of the Hamiltonian 
\begin{eqnarray}
\label{hcd}
\widehat H_{cd}(\vec q)=m_c+m_d+\frac{(\vec k+\vec q)^2}{2m_c}+\frac{\vec k^{\,2}}{2m_d}+V_{cd}(r).   
\end{eqnarray}
such that 
\begin{eqnarray}
\label{en}
\widehat H_{cd}(\vec q)|D_n(-\vec q)\rangle=
E_{D_n}(\vec q) |D_n(-\vec q)\rangle.  
\end{eqnarray}
In this equation $E_{D_n}(\vec q)$ is the 
nonrelativistic energy of the bound state $D_n$ with the 3-momentum $-\vec q$  
\begin {eqnarray}\label{En}
E_{D_n}(\vec q)=M_{D_n}+\frac{\vec q^{\,2}}{2(m_c+m_d)},\qquad  
M_{D_n}=m_c+m_d+\epsilon_{D_n}.  
\end{eqnarray}
The expression (\ref{T1}) for the decay amplitude $T$ now takes the form  
\begin {eqnarray}
\label{Tnr}
T(q^0,\vec q)&=&
\sum_n^\infty |F_n(\vec q)|^2\frac{1}{M_B-E_{D_n}(\vec q)-q^0}, 
\end{eqnarray}
where $F_{n}(\vec q)$ is the $B\to D_n$ transition form factor, 
\begin {eqnarray}
F_{n}(\vec q)=\langle B|J|D_n(-\vec q)\rangle,\qquad M_B=m_b+m_d+\epsilon_B, \nonumber
\end{eqnarray}
and the sum runs over all $c\bar d$ resonances. 
The expression (\ref{Tnr}) can be also written as 
\begin {eqnarray}
\label{Gnr}
T(q^0,\vec q)&=&
\langle B|J\frac{1}{M_B-q^0-\widehat H_{cd}(\vec q)}J^+|B \rangle, 
\end{eqnarray}
where $(\hat H_{cd}(\vec q)-E)^{-1}=G_{cd}(\vec q, E)$ 
is the full off-energy-shell Green function (propagator) of the $c\bar d$ 
system.  
The $B$ decay amplitude is thus given by an
average of the Green function 
$\widehat G_{cd}(\vec q,E)$ at the point $E=M_B-q^0$ over the $B$ meson. 

Let us specify the transition current operator $\hat J_{b\to c}$. 
For the sake of argument we neglect the quark spin effect and
consider spinless nonrelativistic quarks and choose the 
quark current in the form 
\begin{eqnarray}
\label{qcurrent}
\hat J_{b\to c}=\int d\vec k  d\vec k' \hat b(\vec k') \hat c^+(\vec k), 
\end{eqnarray}
where $\hat c(\hat b)$ is the annihilation operator of the $c(b)$ quark
\footnote{Notice that the standard scalar current reads  $\hat J_{b\to c}$=$\int
\frac{d\vec k}{2k^0} \frac{d\vec k'}{2k^{'0}} \hat b(\vec k')\hat c^+(\vec k) $ 
and in the nonrelativistic limit takes the form $\simeq \int {d\vec k} \hat
c^+(\vec k) {d\vec k'} \hat b(\vec k')\left(1-\frac{\vec k^{\,2}}{4m_c^2}\right)
\left(1-\frac{\vec k^{'2}}{4m_b^2}\right)$. Neglecting the factor  
$\left(1-\frac{\vec k^{\,2}}{4m_c^2}\right)\left(1-\frac{\vec
k^{'2}}{4m_b^2}\right)$ as done in  Eq. (\ref{qcurrent}) leads to technical
simplifications both in the OPE  and in the exclusive sum. It can be easily
realized that a particular choice of the current however does not touch any 
arguments related to duality.}.  

For the quark current (\ref{qcurrent}) the $B\to D_n$ transition form factor 
in the rest frame of the $B$-meson reads 
\begin {eqnarray}
\label{current}
F_n(\vec q)=\int d\vec k_q\,\psi_B(\vec k_q) 
\psi_{D_n}\left(\vec k_q+\frac{m_{d}}{m_{c}+m_{d}}\vec q\right),  
\end{eqnarray}
where $\vec k_q$ is the momentum of the light spectator. 

Similarly, for the current (\ref{qcurrent}) the expression (\ref{Gnr}) takes the form 
\begin {eqnarray}
\label{t1}
T(q^0,\vec q)&=&
\langle B|\frac{1}{M_B-q^0-\widehat H_{cd}(\vec q)}|B \rangle.  
\end{eqnarray}

\section{The OPE of the decay rate}
The main idea in constructing the OPE series for $T$, Eq.(\ref{ope}), is to single 
out $\widehat H_{bd}$ from $\widehat H_{cd}(\vec q)$ in the denominator in Eq.~(\ref{t1}) 
and to use the eigenvalue equation (\ref{eigen}). 
First, let us introduce the operator $\delta H(\vec q)$ which measures the
difference of the denominator of Eq.~(\ref{t1}) from the inverse Green function 
of the free-quark transition
\begin{eqnarray}
\label{dh}
M_B-q^0-\widehat H_{cd}(\vec q)&=&
(\delta m-\frac{\vec q^{\,2}}{2m_c}-q^0)-\delta H(\vec q).
\end{eqnarray}
Explicitly, one finds
\begin{eqnarray}
\label{dh1}
\delta H(\vec q)&=&\widehat H_{cd}(\vec q)-m_c-m_d-\frac{\vec q^2}{2m_c}-\epsilon_B.
\end{eqnarray}
Next, isolating $h_{bd}$ in $\delta H(\vec q)$ we obtain 
\begin{eqnarray}
\label{dh2}
\delta H(\vec q)&=&\left(h_{bd}-\epsilon_B\right)
+\frac{1}{2}\left(\frac{1}{m_c}- \frac{1}{m_b}\right)(\vec k^2+V_1)
+\frac{\vec k\cdot\vec q}{m_c}+O\left(\frac{\beta^3\delta m}{m_c^3}\right),   
\end{eqnarray}
where the scale $\beta$ is provided by the hadronic matrix elements 
\begin {eqnarray}\label{beta}
\beta^2\simeq \langle B|\vec k^{\,2}|B \rangle\simeq \langle B|\hat V_1|B \rangle. 
\end{eqnarray}
As already mentioned, 
$\beta$ is of the order of $\Lambda$. The quantity $V_1$ here
is a part of the expansion of the potential $V_{Qq}$ in powers  of $1/m_Q$ 
\begin
{eqnarray} V_{Qq}=V_0+\frac{1}{2m_Q}V_1+\frac{1}{2m_Q^2}V_2+O\left(\frac{\Lambda^4}{m_Q^3}\right). 
\end{eqnarray}

Eqs. (\ref{t1}) and (\ref{dh}) allow us  to construct the expansion of 
$T(q^0,\vec q^{\,2})$ in inverse powers of $\delta m-\frac{\vec q^{\,2}}{2m_c}-q^0$
as follows 
\begin{eqnarray} 
\label{expdh} 
T(q_0,\vec q^2)&=&\frac{1}{\delta m-\frac{\vec q^{\,2}}{2m_c}-q^0}
\sum_{i=0}^{\infty}
\frac{\langle B|\left(\delta H(\vec q)\right)^i|B \rangle}
     {\left(\delta m-\frac{\vec q^{\,2}}{2m_c}-q^0\right)^i}
\end{eqnarray} 
Making use of Eq. (\ref{dh2}) we obtain 
\begin{eqnarray} 
\label{18} 
T(q_0,\vec q^2)&=&\langle B|B \rangle\frac{1}{\delta m-\frac{\vec
q^2}{2m_c}-q^0}+\langle B|(h_{bd}-\epsilon_B)|B \rangle \frac{1}{\left(\delta m-\frac{\vec
q^2}{2m_c}-q^0\right)^{2}}  \nonumber\\ &+&\langle B|\left[\frac{\delta m}{2m_c^2}(\vec
k^2+V_1)-\frac{\vec k\cdot\vec q}{m_c}\right]|B \rangle \frac{1}{\left(\delta m-\frac{\vec
q^2}{2m_c}-q^0\right)^{2}}\nonumber\\
&+&\sum_{i,j=1}^{3}\sum_{n=0}^{\infty}q_iq_j
\langle B|\left[\frac{k_i(h_{bd}-\epsilon_B)^nk_j}{m_c^2}\right]|B \rangle
\frac{1}{\left(\delta m-\frac{\vec q^{\,2}}{2m_c}-q^0\right)^{3+n}}+O(\Lambda^2/m_c^3).
\end{eqnarray} 
The remainder has the order $O(\Lambda^2/m_c^3)$ if we keep 
$\vec q^2\simeq \delta m^2$ and $q_0$ fixed. 

Finally, using Eq. (\ref{eigen}) and the relations $\langle B|k_i|B \rangle=0$ and 
$\langle B|k_ik_j|B \rangle=\frac13\delta_{ij}\langle B|\vec k^{\,2}|B \rangle$ we find  
the following OPE series 
\begin{eqnarray}
\label{19} 
T(q^0,\vec q^{\,2})&=&
\frac{1}{\delta m-\frac{\vec q^{\,2}}{2m_c}-q^0} 
+\frac{\delta m}{2m_c^2}\langle B|\vec k^{\,2}+V_1|B \rangle 
\frac{1}{\left(\delta m-\frac{\vec q^{\,2}}{2m_c}-q^0\right)^{2}}\nonumber\\
&&+\sum_{n=2}^{\infty}\frac{\langle B|\hat O_{n-2}|B \rangle}{3m_c^2} \vec q^{\,2}
\frac{1}{\left(\delta m-\frac{\vec q^{\,2}}{2m_c}-q^0\right)^{n+1}}
+O(\Lambda^2/m_c^3),  
\end{eqnarray} 
where
$\hat O_n=\sum_{j=1}^{3} k_j(h_{bd}-\epsilon_B)^{n}k_j$.  Hereafter the $\sum$ symbol
is omitted. We denote $\langle B|\vec k^{\,2}+V_1|B \rangle=\beta_0^2$, $\beta_0\simeq\Lambda$. 

The series (\ref{19}) is a double expansion of $T(q^0,q^2)$ in $\Lambda/m_c$ and 
$\Lambda/(\delta m-\frac{\vec q^{\,2}}{2m_c}-q^0)$, limited to second order in 
$\Lambda/m_c$ and expanded to all orders in 
$\Lambda/(\delta m-\frac{\vec q^{\,2}}{2m_c}-q^0)$. 
The poles are at
\begin{eqnarray} 
q^0_c(\vec q^{\,2})= \delta m-\frac{\vec q^{\,2}}{2m_c}.  
\end{eqnarray} 
The {\it first term} in (\ref{19}) 
gives {\it the free quark decay amplitude}. 
A remarkable feature of this series is that the $\Lambda/\delta m$
and $\Lambda/m_c$ corrections to the 
free-quark decay are absent thanks to Eq.~(\ref{eigen}) and the relation
$\langle B|k_i|B \rangle=0$.
The expansion (\ref{19}) substitutes the whole set of 
hadron poles by a complicated 
quark singularity at the point $q^0=q^0_c(\vec q^{\,2})$. 

Let us treat the series (\ref{19}) {\it formally} and calculate the 
integrated rate which is 
obtained as a double expansion in $\Lambda/m_c$ and $\Lambda/\delta m$.

Let us rewrite the expression (\ref{19}) as follows 
\begin{eqnarray}
\label{20}
T(q^0,\vec q^{\,2})=\left(1-\frac{\langle B|\vec k^{\,2}+V_1|B \rangle}{2m_c^2}
\delta m\frac{\partial}{\partial \delta m}+\vec q^{\,2}
\sum_{n=2}^{\infty}\frac{\langle B|\hat O_{n-2}|B \rangle}{3m_c^2}\frac{1}{n!}
\left(-\frac{\partial}{\partial \delta m}\right)^n \right) 
\frac{1}{\left(\delta m-\frac{\vec q^{\,2}}{2m_c}-q^0\right)}
\end{eqnarray} 
This representation is very convenient for the calculation of the decay rate 
Eq. (\ref{ratecontour}): the integration over $q^0$ is now easily performed since 
\begin{eqnarray}
\int\limits_{C(\vec q^{\,2})} dq^0 L((q^0)^2-\vec q^{\,2})
\theta(q^0>|\vec q|)\frac{1}{\left(\delta m-\frac{\vec q^{\,2}}{2m_c}-q^0\right)}=
L(q^0_c(\vec q^{\,2}),\vec q^{\,2})
\theta \left(|\vec q|<-m_c+\sqrt{m_c^2+2m_c\delta m}\right),  
\end{eqnarray} 
where the $\theta$-function 
$\theta \left(|\vec q|<-m_c+\sqrt{m_c^2+2m_c\delta m}\right)$ 
reflects the fact that the left crossing of the contour with the real axis 
in the complex $q^0$ plane should always happen at the point 
${\rm Re} (q^0)=|\vec q|$.   
The integrated rate is then given by the expression\footnote{
A remark is in order here. When computing the decay rate in Eq.(\ref{47}) we have
 interchanged the derivation with respect to $\delta m$ and the integration 
 over $dq^0$. 
 We can also directly integrate the expression Eq. (\ref{19}) over $dq^0$.   
 In this case we should take into account that the contour 
 $C(\vec q^{\,2})$ in the complex $q^0$-plane always lies on the r.h.s. of the line 
 ${\rm Re}(q^0)=|\vec q|$. If we erroneously do not take this condition into account 
 then multiple poles in (\ref{19}) do not contribute at all since the complex integral 
 vanishes when the multiple poles are inside the contour as well as when they are 
 outside (see also discussion in \cite{bsuv93}). 
 To proceed correctly, one should replace a multiple pole, say a double pole, by an equivalent 
 set of two neighbouring poles.  
 Then the crossing of the boarder gives a non-vanishing result when one of the two poles 
 is inside the contour and the other one is outside. 
 Integrating before taking the derivative with respect to $\delta m$ as in Eq. (\ref{47}) 
 corresponds to treating in a specific way the crossing of the boarder by the multiple poles. 
 We show elsewhere \cite{15r} that both treatments lead to the same result.}   
\begin{eqnarray}
\label{47}
\Gamma^{OPE}(B\to X_cl\nu)=\left(1-\frac{\langle B|\vec k^{\,2}+V_1|B \rangle}{2m_c^2}
\delta m\frac{\partial}{\partial \delta m} \right)I_1(\delta m, m_c)
+\sum_{n=2}^{\infty}\frac{\langle B|\hat O_{n-2}|B \rangle}{3m_c^2}\frac{1}{n!}
\left(-\frac{\partial}{\partial \delta m}\right)^n  I_3(\delta m, m_c),  
\end{eqnarray}
where 
\begin{eqnarray}
I_n(\delta m, m_c)=\int\limits^{-m_c+\sqrt{m_c^2+2m_c\delta m}}_{0}
d\vec q^{\,2} |\vec q|^n L\left ( [q_c^0(\vec q^{\,2})]^2 -\vec q^{\,2} \right ) . 
\end{eqnarray}  
For the free quark decay one finds
\begin{eqnarray}
\Gamma(b\to cl\nu)=I_1(\delta m, m_c). 
\end{eqnarray}
Let us consider the leptonic tensor of the general form $L(q^2)=(q^2)^N$. 
For semileptonic decays to massless spin 1/2 leptons $N=1$. 
The case $N=0$ corresponds to scalar leptons.
Since the leptonic tensor is proportional to $q^2$, 
it is now convenient to introduce a new integration variable 
$q^2$ as follows 
\begin{eqnarray}
q^2=\left [ q^0_c(\vec q^{\,2}) \right ]^2 -\vec q^{\,2}.
\end{eqnarray}
Then the integrated rate takes the form 
\begin{eqnarray}
\label{31e}
\Gamma^{OPE}(B\to X_cl\nu)&=&\Gamma(b\to cl\nu) \\
&-&\frac{\langle B|\vec k^{\,2}+V_1|B \rangle}{2m_c^2}
\delta m\frac{\partial}{\partial \delta m}I_1(\delta m,m_c)
+\sum_{n=2}^{\infty}\frac{\langle B|\hat O_{n-2}|B \rangle}{3m_c^2}\frac{1}{n!}
\left(-\frac{\partial}{\partial \delta m}\right)^n
I_3(\delta m,m_c), 
\nonumber
\end{eqnarray}
where we have taken into account that $I_1(\delta m, m_c)$ gives the exact 
free-quark decay rate. 
A simple algebraic exercise gives to the $1/m_c^2$ accuracy
\begin{eqnarray}
I_1(\delta m, m_c)&=&(\delta m)^{2N+3}\left[ A^N_{1/2}\left( 1-\frac{3}{2}
\frac{\delta m}{m_c}
+\frac{15}{8}\frac{\delta m^2}{m_c^2}\right)
+\frac{5}{8}A^N_{3/2}\frac{\delta m^2}{m_c^2}
+O\left(\frac{\delta m^3}{m_c^3}\right)\right],
\nonumber\\
I_3(\delta m, m_c)&=&(\delta m)^{2N+5}\left[A^N_{3/2}
+O\left(\frac{\delta m}{m_c}\right)\right]
\end{eqnarray}
where 
\begin{eqnarray}
\label{a}
A^N_{m}=\int_0^1 dx x^m(1-x)^N=B(N+1,m+1), 
\qquad A^N_{3/2}=\frac{3}{2N+5}A^N_{1/2}, 
\end{eqnarray}
$B(p,q)$ being the Euler function. Finally, we come to the relation  
\begin{eqnarray}
\label{finalope}
\frac{\Gamma(B\to X_cl\nu)}{\Gamma(b\to cl\nu)}&=&1+
\frac{\langle B|\vec k^{\,2}|B \rangle}{2m_c^2}-(2N+3)\frac{\langle B|V_1|B \rangle}{2m_c^2}
+\sum\limits_{n=3}^{2N+5}\frac{(-1)^n\,C^n_{2N+5}}{2N+5}\,\frac{\langle B|\hat O_{n-2}|B \rangle}
{m_c^2\delta m^{n-2}}+O\left(\frac{\Lambda^2\delta m}{m_c^3}\right), 
\end{eqnarray}
with $C^k_{n}=\frac{n!}{k!(n-k)!}$. Notice that the coefficient of the term 
$\langle B|\vec k^{\,2}|B \rangle$ 
does not depend on $N$, i.e. it does not depend on the form of the leptonic tensor. 
 
Summing up, the OPE predicts the following features of the inclusive SL decay rate:
\begin{itemize}
\item
The LO term  reproduces the rate of the free-quark decay process $b\to c$. 
\item 
The $1/m_c$ and $1/\delta m$ corrections are absent. This is due to the fact that 
the average over the $B$-state of the operator of the relevant dimension vanishes. 
\item 
Lowest-order corrections to the free-quark process emerge in the $1/m_c^2$ order.  
A main part of these corrections is due to the average values of the 
dimension-2 operators  ${\langle B|\vec k^{\,2}|B\rangle}$ and 
${\langle B|\vec k^{\,2}+V_1|B\rangle}$. Also
the  operators $\hat O_n=k_j(h_{bd}-\epsilon_B)^nk_j$ contribute in the $1/m_c^2$ order.
Their contribution is however suppressed with the additional powers of $\delta m$.
\end{itemize}
In the next section we shall
analyze the accuracy of the OPE predictions. 

\section{Heavy quark expansion and the hadronic sum rules}
Before proceeding with the direct summation of the exclusive channels one by one we 
derive hadronic sum rules which are important for the comparison of the 
exact result with the OPE analysis. 

\subsection{Heavy quark expansion of the form factors in the potential model}

The wave function of the $Q\bar q$ bound state has the form 
\begin{eqnarray}
\Psi_{\vec p}(\vec k_Q, \vec k_q)&=&\delta(\vec p-\vec k_Q-\vec k_q)
\psi\left(\frac{m_Qm_q}{m_Q+m_q}\left(\frac{\vec k_q}{m_q}-\frac{\vec k_Q}{m_Q}\right)\right)\nonumber\\
&=&\delta(\vec p-\vec k_Q-\vec k_q)\psi\left(\vec k_q-\frac{m_q}{m_Q+m_q}\vec p\right) 
\end{eqnarray}
where $\vec p$ is the momentum of the bound state. 

The $B\to D_n$ transition form factor is the average over the meson states of the operator 
$\Omega_{bc}(\vec q)$ given by the following kernel 
\begin{eqnarray}
\langle \vec k_b|\Omega_{bc}(\vec q)|\vec k_c \rangle=\delta(\vec k_b-\vec k_c-\vec q). 
\end{eqnarray}
So the transition form factor is defined by the following expression 
\begin{eqnarray}
\langle B(\vec p_B)|\Omega_{bc}(\vec q)|D_n(\vec p_n)\rangle=
\delta(\vec p_B-\vec p_n-\vec q)
F_n\left((\vec v_B-\vec v_n)^2 \right), 
\end{eqnarray}
with $\vec v_B=\vec p_B/(m_b+m_q)$ and $\vec v_n=\vec p_n/(m_c+m_q)$ and 
\begin{eqnarray}
F_n\left(\vec v_B-\vec v_n\right)=\int d\vec k_q 
\psi_B\left(\vec k_q-\frac{m_q}{m_b+m_q}\vec p_B\right) 
\psi_{D_n}\left(\vec k_q-\frac{m_q}{m_c+m_q}\vec p_n\right).  
\end{eqnarray}
A simple change of variables $\vec k_q\to \vec k_q + m_q/(m_b+m_q)\vec p_B$
makes  it obvious that  the decay form factor depends on the square of the
relative 3-velocities  of the initial and final mesons, and not on the relative
3-momentum squared as the   elastic form factor.  Nevertheless in the $B$ rest
frame we write 
\begin{eqnarray}
\label{ff}
F_n(\vec q)=\int d\vec k_q 
\psi_B(\vec k_q) 
\psi_{D_n}\left(\vec k_q+\frac{m_q}{m_c+m_q}\vec q\right).  
\end{eqnarray}

The wave function $\psi_{Q\bar q}$ is an eigenstate of the Hamiltonian 
\begin{eqnarray}
h_{Q\bar q}=\frac{\vec k^{\,2}}2\left(\frac  1{m_Q}+\frac 1{m_q}\right)+V_{Qq}(r), 
\end{eqnarray}
where $\vec k$ is the conjugate variable to $\vec r$. 
The transition form factors $F_n$ has some general properties independent of the details of the 
potential $V_{Q\bar q}$. 
Such properties of the 
transition form factors are derived by performing the HQ expansion of the Hamiltonian. 
To this end we 
apply the usual Quantum Mechanical perturbation theory. 

For our purposes it is convenient to consider $h_{bd}$ as the full hamiltonian, 
$h_{cd}$ as a nonperturbed Hamiltonian, 
and $\widehat{U}=h_{bd}-h_{cd}$ as the perturbation.  
The perturbation has the form 
\begin{eqnarray}
\hat U=\frac{1}{2}\left(\frac{1}{m_b}-\frac{1}{m_c}\right)(\vec k^{\,2}+V_1)
+O\left(\beta^3\frac{\delta m}{m_c^3}\right)
=-\frac{\delta m}{2m_c^2}(\vec k^{\,2}+V_1)+O\left(\beta^2\frac{\delta m^2}{m_c^3}\right)
\end{eqnarray}
where we assume the following expansion of the $Q\bar q$ potential 
\begin{eqnarray}
V_{Q\bar q}=V_0+\frac{1}{2m_Q}V_1+\frac{1}{2m_Q^2}V_2+...
\end{eqnarray}
The perturbation has the order $\delta m/m_c^2$ such that 
we can construct the HQ expansion of the wave functions and the binding energies. 
Let us remind the standard formulas:
Let $\{\psi_{D_n}\}$ be the full system of eigenstates of the $h_{cd}$, 
and the $\{\epsilon_{D_n}\}$ the corresponding eigenvalues. 
Then, the mass of the $n$-th excitation in the $c\bar d$ system reads 
$M_{D_n}=m_c+m_d+\epsilon_{D_n}$. 
Let $\{\psi_{B_n}\}$ be the full system of eigenstates of the $h_{bd}$, and the 
$\{\epsilon_{B_n}\}$ 
the corresponding eigenvalues. 

The standard formulas give 
\begin{eqnarray}
\psi_{B_n}=\psi_{D_n}+\sum_{m\ne n}\frac{U_{mn}}{\epsilon_{D_n}-\epsilon_{D_m}}\psi_{D_m}
+O(\delta m^2\beta^2/m_c^4)
\end{eqnarray}
and 
\begin{eqnarray}
\label{e}
\epsilon_{B_n}=\epsilon_{D_n}+U_{nn}+\sum_{m\ne n}
\frac{|U_{mn}|^2}{\epsilon_{D_n}-\epsilon_{D_m}}+...
\end{eqnarray}
where \footnote{$\epsilon_{B_0}$ is just 
$\epsilon_B$ defined in Eq.~(\ref{epsb}). 
We also usually write $B$ instead of $B_0$.}
\begin{eqnarray}\label{umn}
U_{mn}=-\frac{\delta m}{2m_c^2}\langle \psi_{D_n}|\vec k^{\,2}+V_1|\psi_{D_m}\rangle+
O\left(\beta^2\frac{\delta m^2}{m_c^3}\right) 
\end{eqnarray}
The excitation energies satisfy the relation 
\begin{eqnarray}
\epsilon_{D_n}-\epsilon_{D_m}\simeq (n-m)\lambda_{nm}, \quad \lambda_{mn}\simeq \beta.  
\end{eqnarray}
In terms of the wave functions, the transition form factor (\ref{ff}) 
takes a simple form 
\begin{eqnarray}
F_n(\vec q)=\langle \psi_{B_0}|\psi_{D_n}(-\vec q)\rangle 
\end{eqnarray}

The expansion of the wave function $\psi_{B_0}\equiv \psi_{B}$ reads 
\begin{eqnarray}
\psi_{B_0}=\psi_{D_0}+\sum_{m\ne 0}
\frac{1}{2}\left(\frac{1}{m_b}-\frac{1}{m_c}\right)
\frac{\langle \psi_{D_0}|\vec k^{\,2}+V_1|\psi_{D_m}\rangle }
{\epsilon_{D_n}-\epsilon_{D_m}}\psi_{D_m}+O(\delta m^2\beta^2/m_c^4),  
\end{eqnarray}
such that 
\begin{eqnarray}
\label{ffbd}
F_n(\vec q)&=&\langle \psi_{D_0}
+\sum_{m\ne 0}\frac{1}{2}\left(\frac{1}{m_b}-\frac{1}{m_c}\right)
\frac{\langle \psi_{D_0}|\vec k^{\,2}+V_1|\psi_{D_m}\rangle^* }
{\epsilon_{D_0}-\epsilon_{D_m}}
\psi_{D_m}|\psi_{D_n}(\vec q)\rangle
+O(\delta m^2\beta^2/m_c^4)\nonumber\\
&=&f_{0n}(\vec q)
+\sum_{m\ne 0}\left(-\frac{\delta m}{2m_c^2}\right)
\frac{\langle \psi_{D_0}|\vec k^{\,2}+V_1|\psi_{D_m}\rangle^* }
{\epsilon_{D_0}-\epsilon_{D_m}}
f_{mn}(\vec q)
+O(\delta m^2\beta^2/m_c^4)
\end{eqnarray}
where $f_{mn}(\vec q)=\langle \psi_{D_m}|\psi_{D_n}(\vec q)\rangle$. 
By virtue of (\ref{ff}) one obtains
\begin{eqnarray}
\label{r}
f^2_{nn}(\vec q)&=&1-r^2_{nn}\frac{\vec q^{\,2}}{m_c^2}+O(\vec q^{\,4}/m_c^4),\nonumber \\
f^2_{nm}(\vec q)&=&\quad\;\;\, r^2_{nm}\frac{\vec q^{\,2}}{m_c^2}+O(\vec q^{\,4}/m_c^4),
\qquad m\ne n,
\end{eqnarray}
with $r_{mn}$ being numbers of order unity plus higher order $1/m_c$ corrections. 
We shall use the notation $r_{n}=r_{n0}$. 
Notice that the radii $r_n$ describe the form factors of the transitions between 
different levels in the $c\bar d$ system ($D_0\to D_n$) and so know 
nothing about $\delta m$.  

We now rewrite Eq. (\ref{ffbd}) as follows   
\begin{eqnarray}
\label{rho}
F_0(\vec q)&=&1-r^2_{0}\frac{\vec q^{\,2}}{2m_c^2} 
+\sum_{m\ne 0}
\left(-\frac{\delta m}{2m_c^2}\right)
\frac{\langle \psi_{D_0}|\vec k^{\,2}+V_1|\psi_{D_m}\rangle^* }{\epsilon_{D_0}-\epsilon_{D_m}}
f_{m0}(\vec q)
+O\left(\frac{\delta m^2\beta^2}{m_c^4}\right). 
\end{eqnarray}

At $\vec q^{\,2}=0$ we thus come to the relation
\begin{eqnarray}
\label{nrluke}
F_0(0)&=&1+O(\delta m^2\beta^2/m_c^4).  
\end{eqnarray}
One can see that the $O(\delta m\beta/m_c^2)$ term in $F_n(0)$ is absent. 
This is a non-relativistic analog of the Luke theorem \cite{luke}. 

For the squares of the form factors we obtain the following important 
relations\footnote{
At any $n$ states with angular momenta $L=0,...,n$ exist. The form factors  
$f_n(\vec q^{\,2})$ and $F_{n}(\vec q^{\,2})$ are thus understood as properly normalized sums 
$\sum_{L=0}^n f_{nL}(\vec q)$ and $\sum_{L=0}^n F_{nL}(\vec q)$, respectively.}   
\begin{eqnarray}
\label{result5}
F_0^2(\vec q)&=&1-\rho_0^2\frac{\vec q^{\,2}}{m_c^2}+
O\left(\frac{\delta m^2\beta^2}{m_c^4}\right),\qquad 
\rho_0^2=r_0^2+O\left(\frac{\beta\delta m}{m_c^2}\right); 
\nonumber\\
F_n^2(\vec q)&=&\quad\;\;\, \rho_n^2\frac{\vec q^{\,2}}{m_c^2}+
O\left(\frac{\delta m^2\beta^2}{m_c^4}\right),\qquad 
\rho_n^2=r_n^2+O\left(\frac{\beta\delta m}{m_c^2}\right). 
\end{eqnarray}
As we shall see later the radii $\rho_n$ (as well as $r_n$) are not independent and satisfy 
certain sum rules. The relations (\ref{result5}) are the main result 
of this section. They are necessary for the calculation of the decay rates.
 
\subsection{Inclusive hadronic sum rules}

To obtain a non-relativistic equivalent of the whole tower of sum rules
\cite{greg}, i.e. the Bjorken sum rule, the Voloshin and the higher moments,
we consider the following set of quantities ($i=0,1,\dots$) 
\begin{eqnarray}
\label{siq} 
S_i(\vec q)=\langle B| (\delta H(\vec q))^i|B\rangle, 
\end{eqnarray}
where $\delta H(\vec q)$ is defined in Eq. (\ref{dh}). 
Notice that $S_i(\vec q)$ appear in the expansion for $T(q^0,\vec q^2)$, 
Eq. (\ref{expdh}).  
We shall derive two different representations for $S_i(\vec q)$ 
and obtain sum rules equating these representations. 

The first representation is obtained by inserting the full system of the 
eigenstates $|D_n(-\vec q)\rangle$ of the Hamiltonian $H_{cd}(\vec q)$ in Eq. (\ref{siq}). 
The $|D_n(-\vec q)\rangle$ are also eigenstates of the operator 
$\delta H(\vec q)$ that is made obvious using 
$\delta H_{cd}(\vec q)$ in the form Eq. (\ref{dh1}): 
\begin{eqnarray}
\delta H(\vec q)|D_n(-\vec q)\rangle=\delta_n(\vec q)|D_n(-\vec q)\rangle,
\qquad \delta_n(\vec q)=\epsilon_{D_n}-\epsilon_B+\frac{\vec q^2}{2(m_c+m_d)}
-\frac{\vec q^2}{2m_c}. 
\end{eqnarray}
As a result of inserting the full system we find  
\begin{eqnarray}
\label{si1}
S_i(\vec q)=\sum_{n=1}^{\infty}|F_n(\vec q)|^2\left(\delta_n(\vec q)\right)^i.
\end{eqnarray}
Eq. (\ref{e}) gives the following expansion for $\delta_n(\vec q)$: 
\begin {eqnarray}
\label{delta}
\delta_n(\vec q)&=&
\Delta_n+\frac{1}{2}\left(\frac{1}{m_c}-\frac{1}{m_b}\right)
\langle {D_0}|\vec k^{\,2}+V_1|{D_0}\rangle-
\frac{m_d\vec q^{\,2}}{2m_c(m_c+m_d)}+O\left(\frac{\delta m^2\beta^3}{m_c^4}\right),
\end{eqnarray}
where 
\begin{equation}
\label{deln}
\Delta_n\equiv \epsilon_{D_n}-\epsilon_{D_0}
\end{equation}
Notice that within the leading-order accuracy we can replace 
$\langle D_0|\vec k^{\,2}+V_1|D_0\rangle$ with $\langle B|\vec k^{\,2}+V_1|B \rangle$.

Another representation for $S_i(\vec q)$ is obtained by using $\delta H(\vec q)$ 
in the form (\ref{dh2}): 
\begin {eqnarray}
\label{si}
S_i(\vec q)=\langle B|\left(\delta H(\vec q)\right)^i|B \rangle=
\langle B| \left(\hat h_{bd}-\epsilon_B+
\frac{\vec k^2+V_1}{2}\left(\frac1m_c-\frac1m_b\right) 
+\frac{\vec k\cdot\vec q}{m_c}+O\left(\frac{\beta^3\delta m}{m_c^3}\right)
\right)^i|B \rangle. 
\end{eqnarray}
This formula gives $S_i(\vec q)$ in terms of the matrix elements of various operators over
the $B$-meson. 

The representations (\ref{si1}) and (\ref{si}) for $S_i(\vec q)$ 
provide the l.h.s. and the r.h.s. of the sum rules, respectively. 
Let us notice that terms denoted by $O(\beta^3\delta m/m_c^3)$ 
in Eqs. (\ref{delta}) and (\ref{si}) do not depend on $\vec q$. All 
$\vec q$-dependent terms are shown explicitly.  

Using Eq. (\ref{eigen}) and the relations  
$\langle B|k_i|B \rangle=0$ and 
$\langle B|k_ik_j|B \rangle= {1 \over 3} \delta_{ij}\langle B|\vec k^{\,2}|B \rangle$
we come to the set of sum rules. In fact each of these sum rules is equivalent to an
infinite number of relations at different powers of $\vec q^2$ and $1/m_c$. 

i=0:  
\begin{eqnarray}
\label{sr0}
S_0&=&\sum\limits_n^\infty |F_n(\vec q)|^2=1. 
\end{eqnarray}
Obviously the r.h.s. does not depend on $\vec q$. 
At $\vec q^{\,2}=0$ this is an identity. 
Using the definition~(\ref{result5}) and 
comparing the term linear in $\vec q^{\,2}$ 
we find the {\it NR Bjorken sum rule} \cite{13r} 
\begin{eqnarray}
\label{sr0b}
\rho_0^2=\sum_{n=1}^{\infty}\rho_n^2. 
\end{eqnarray}
 
i=1:
The r.h.s. of this sum rule reads    
\begin{eqnarray} \label{sr1}
S_1&=&\frac{1}{2}\left(\frac{1}{m_c}-\frac{1}{m_b}\right)\langle B|\vec k^{\,2}+V_1|B \rangle+
O\left(\frac{\beta^3\delta m}{m_c^3}\right). \end{eqnarray}
where we have used Eqs.~(\ref{epsb}) and (\ref{hbd}). The r.h.s. of this 
sum rule is also independent of $\vec q$. 
From the definition
(\ref{siq}) and using (\ref{delta}) as well as the SR (\ref{sr0}), 
we rewrite the Eq.~(\ref{sr1})  as follows 
\begin{eqnarray} 
\label{sr1a} 
\sum_{n=0}^{\infty}F_n^2(\vec q)\Delta_n=
\frac{\vec q{\,^2} m_d}{2m_c(m_c+m_d)} 
+O\left(\frac{\beta^3\delta m}{m_c^3}\right). 
\end{eqnarray}
Notice that the terms $O(\beta^2 \delta m/m_c^2)$ cancel between r.h.s amd l.h.s.
Comparing the linear in $\vec q^{\,2}$ term 
yields the {\it NR Voloshin sum rule} \cite{14r} 
\begin{eqnarray}
\label{sr1b}
\sum_{n=1}^{\infty}\rho_n^2\Delta_n=\frac{m_d}{2}\frac{1}{1+{m_d \over m_c}}.  
\end{eqnarray}
Let us notice that the r.h.s. of the Eq. (\ref{sr1b}) does not contain 
higher-order $1/m_c$ corrections. 

Combining the Bjorken and the Voloshin sum rules provides a simple 
constraint on the parameter $\rho_0^2$ which is in fact the slope of the Isgur-Wise function. 
Namely, 
\begin{eqnarray}
\rho_0^2=\sum_{n=1}^{\infty}\rho_n^2=\frac{1}{\Delta_1}\sum_{n=1}^{\infty}\rho_n^2\Delta_1
<\frac{1}{\Delta_1}\sum_{n=1}^{\infty}\rho_n^2\Delta_n=\frac{m_d}{2\Delta_1}\frac{1}{1+m_d/m_c}
<\frac{m_d}{2\Delta_1},
\end{eqnarray}
where $\Delta_n$ are defined in Eq. (\ref{deln})

i=2: 
The r.h.s. of this sum rule reads  
\begin{eqnarray}
\label{sr2}
S_2&=&\frac{\vec q^{\,2}\langle B|\vec k^{\,2}|B \rangle}{3m_c^2}
+O\left(\frac{\beta^4\delta m^2}{m_c^4}\right).
\end{eqnarray}

Using (\ref{sr0}) and (\ref{sr1}) yields for the l.h.s. 
\begin{eqnarray}
\label{sr2a}
\sum_{n=0}^{\infty}F_n^2(\vec q)\Delta_n^2=
\frac{\vec q^{\,2}\langle B|\vec k^2|B \rangle}{3m_c^2}
\left(1+O\left(\frac{\beta^2\delta m}{m_c^3}\right)\right)
+\frac{\vec q^{\,4}m_d^2}{4m_c^2(m_c+m_d)^2}
+O\left(\frac{\beta^4\delta m^2}{m_c^4}\right). 
\end{eqnarray}
The linear $\vec q^{\,2}$-term yields 
\begin{eqnarray}
\label{sr2b}
\sum_{n=1}^{\infty}\rho_n^2\Delta_n^2=\frac{1}{3}\langle B|\vec k^2|B \rangle
\left(1+O\left(\frac{\beta^2\delta m}{m_c^3}\right)\right)
\end{eqnarray}

i$\ge$3: 

For i=3 we find for the r.h.s. 
\begin{eqnarray}
\label{sr3}
S_3&=&\frac{1}{3}\frac{\vec q^{\,2}\langle B|k_j(h_{bd}-\epsilon_B)k_j|B \rangle}{m_c^2}
\left(1+O\left(\frac{\beta\delta m}{m_c^2}\right)\right)
+O\left(\frac{\beta^6\delta m^3}{m_c^6}\right).
\end{eqnarray}

Using (\ref{sr0})-(\ref{sr2}) yields for the l.h.s. 
\begin{eqnarray}
\label{sr3a}
\sum_{n=0}^{\infty}F_n^2(\vec q)\Delta_n^3
&=&\frac{\vec q^{\,2}\langle B|k_j(h_{bd}-\epsilon_B)k_j|B\rangle}{3m_c^2}
\left(1+O\left(\frac{\beta\delta m}{m_c^2}\right)\right)\nonumber\\
&&+\frac{\vec q^{\,4}}{m_c^4}O\left(\frac{m_d^2\beta^2\delta m}{m_c^2}\right)
+\frac{\vec q^{\,6}}{m_c^6}O\left(m_d^3\right)
+O\left(\frac{\beta^5\delta m^2}{m_c^4}\right).
\end{eqnarray}

The linear $\vec q^{\,2}$-term yields 
\begin{eqnarray}
\label{sr3b}
\sum_{n=1}^{\infty}\rho_n^2\Delta_n^3=
\frac{\langle B|k_j(h_{bd}-\epsilon_B)k_j|B \rangle}{3}
\left(1+O\left(\frac{\beta\delta m}{m_c^2}\right)\right)
=\frac{1}{3}\langle B|\hat O_1|B\rangle\left(1+O\left(\frac{\beta\delta m}{m_c^2}\right)\right).
\end{eqnarray}

Similarly at higher $i\ge 3$ one obtains at the $\beta\delta m/m_c^2$
accuracy 
\begin{eqnarray}
\label{sr4b}
\sum_{n=1}^{\infty}\rho_n^2\Delta_n^i=
\frac{1}{3}\langle B|k_j(h_{bd}-\epsilon_B)^{i-2}k_j|B \rangle=\frac{1}{3}\langle B|\hat O_{i-2}|B \rangle.
\end{eqnarray}
These sum rules are used in the next section for comparison of the exact decay rate 
with the OPE result and for analyzing the duality-violation effects. 

\section{Summation over the exclusive channels}

We now proceed to the summation of the exclusive channels. As the first step, let us show that
there is an explicit difference between the exclusive sum and the OPE series.

\subsection{The origin of duality violation}

Proceeding with the sum over the exclusive channels we write: 
\begin {eqnarray}
\label{exclusive_sum}
\Gamma(B\to X_c l\nu)&=&\frac{1}{2\pi i} 
\sum_n^\infty\int dq^2 L(q^2)\int_{C(q^2)}dq^0|\vec q|
\frac{|F_n(\vec q)|^2}{M_b-q^0-E_n(\vec q)}\nonumber \\ 
&=&\frac{1}{2\pi i}\sum_n^\infty\int dq^2 L(q^2)\int_{C(q^2)}dq^0|\vec q|
\frac{|F_n(\vec q)|^2}{\delta m-
\frac{\vec q^{\,2}}{2m_c}-q^0+\delta_n(\vec q)}\nonumber \\
&=&\frac{1}{2\pi i}\sum_n^\infty\int dq^2 L(q^2)\int_{C(q^2)}dq^0|\vec q|
\frac{|F_n(\vec q)|^2}{\delta m-\frac{\vec q^{\,2}}{2m_c}-q^0}
\left[1-\frac{\delta_n(\vec q)}{\delta m-
\frac{\vec q^{\,2}}{2m_c}-q^0}+...\right]\nonumber \\    
&=&\frac{1}{2\pi i}\int dq^2 L(q^2)\sum_n^{n(q^2)}\int_{C(q^2)}dq^0|\vec q|
\frac{|F_n(\vec q)|^2}{\delta m-
\frac{\vec q^{\,2}}{2m_c}-q^0}\left[1-\frac{\delta_n(\vec q)}{\delta m-
\frac{\vec q^{\,2}}{2m_c}-q^0}+...\right],    
\end{eqnarray}
where in the r.h.s. $|\vec q|=\sqrt{(q^0)^2-q^2}$. 
Notice that the sum is truncated at the proper 
$n(q^2)$ which is the maximal number of hadron resonances kinematically allowed at a 
given value of $q^2$, i.e. resonances satisfying the relation 
$M_n<M_B-\sqrt{q^2}$. The contour $C(q^2)$ is responsible for this selection, 
since only the resonances enclosed by the contour contribute into the sum. 
All states which are beyond this contour do 
not contribute. 

Finally, the series (\ref{exclusive_sum}) can be written in the form 
\begin{eqnarray}
\label{77}
\Gamma(B\to X_c l \nu)&=&\int dq^2 L(q^2)\theta(q^2) 
\int dq^0d\vec q^{\,2} |\vec q|\delta((q^0)^2-q^2-\vec q^{\,2}) 
\sum_n^{n(q^2)}|F_n(\vec q)|^2\nonumber\\
&&\left(1+\delta_n(\vec q^{\,2})\frac{\partial}{\partial \delta m}
+\frac{1}{2}\delta_n^2(\vec q^{\,2})\frac{\partial^2}{\partial \delta m^2}+... \right)
\delta(q^0-\frac{\vec q^{\,2}}{2m_c}-\delta m). 
\end{eqnarray}
On the other hand, the sum rules (\ref{sr1})-(\ref{sr3}) allow us to rewrite 
the decay rate (\ref{47}) in the form 
\begin{eqnarray}
\label{78}
\Gamma^{OPE}(B\to X_c l \nu)&=&\int dq^2 L(q^2)\theta(q^2) 
\int dq^0 d\vec q^{\,2} |\vec q|\delta((q^0)^2-q^2-\vec q^{\,2}) 
\sum_n^{\infty}|F_n(\vec q)|^2\nonumber\\
&&\left(1+\delta_n(\vec q^{\,2})\frac{\partial}{\partial \delta m}
+\frac{1}{2}\delta^2_n(\vec q^{\,2})\frac{\partial^2}{\partial \delta m^2}+... \right)
\delta(q^0-\frac{\vec q^{\,2}}{2m_c}-\delta m). 
\end{eqnarray}
It is easy to see that the exact result and the result of the OPE are different 
due to contributions of highly-excited states: {\it at any $q^2$ the OPE picks up
also   the contribution of the resonances forbidden kinematically at this
$q^2$}. Thus the accuracy of duality is determined by the accuracy of violating
the sum rules  connected with the truncation of the exclusive sum, and is 
therefore connected with the  convergence of these sums.  

\subsection{Sum of the exclusive channels and the accuracy of the OPE}
We now calculate the individual decay rates keeping terms of order 
$\frac{\Lambda^2}{m_c^2}\left(\frac{\delta m}{\Lambda}\right)^n$ in the decay rates
but neglecting higher orders 
$\frac{\Lambda^3}{m_c^3}\left(\frac{\delta m}{\Lambda}\right)^n$. 
The necessary expressions with the relevant accuracy are given below.  

$|\vec q|$ in the free-quark decay $b\to c l \nu$ at $q^2$ has the form 
\begin{eqnarray}
|\vec q|=\sqrt{\delta m^2-q^2}
\left(1-\frac{\delta m}{2m_c}+\frac{3}{8}\frac{\delta m^2}{m_c^2}+\frac{\delta
m^2-q^2}{8m_c^2}
\right).
\end{eqnarray}

The general expression for $|\vec q|$ in the $B\to D_n l \nu$ transition at $q^2$ reads
\begin{eqnarray}
|\vec q|_{n}=\sqrt{\delta M_n^2-q^2}
\left(1-\frac{\delta M_n}{2(m_c+m_d)}
+\frac{3}{8}\frac{\delta M_n^2}{(m_c+m_d^2}+\frac{\delta
M_n^2-q^2}{8(m_c+m_d)^2}\right)
\end{eqnarray}
where $\delta M_n=M_B-M_n \simeq \delta m- \Delta_n - \delta m\beta_0/(2 m_c)$
from Eqs.~(\ref{umn}) and (\ref{deln}). 

The necessary accuracy for the transition into the ground state is
\begin{eqnarray}
|\vec q|_{n=0}=\sqrt{\delta m^2\left(1-\frac{\beta_0^2}{2m_c^2}\right)^2-q^2}
\left(1-\frac{\delta m}{2(m_c+m_d)}+\frac{3}{8} \frac{\delta m^2}{m_c^2}
+\frac{\delta m^2-q^2}{8m_c^2}\right).
\end{eqnarray}
Recall that $\beta_0^2=\langle\vec k^2+V_1\rangle$.

For $|\vec q|_{n\ne 0}$ less accuracy is enough since the contribution of the 
$D_n, n\ne 0$ into the SL decay rate is suppressed by the additional factor $\vec q^{\,2}/m_c^2$: 
\begin{eqnarray}
|\vec q|_{n\ne 0}=\sqrt{\left(\delta m-\Delta_n\right)^2-q^2}.
\end{eqnarray}

With these formulas for $|\vec q|$ the decay rates of the exclusive 
channels take the following form.

Free quark decay $b \to c l\nu$:
\begin{eqnarray} {1 \over L(q^2)} {d\Gamma (b \to c l \nu) \over dq^2} &=& |\vec{q}|
\left ( 1 - {\delta m \over m_c} + {3 \over 2} {\delta m^2 \over m_c^2} - {q^2
\over 2 m_c^2} \right )\nonumber \\ &=& \sqrt{\delta m^2 - q^2} \left ( 1 -{3
\over 2} {\delta m \over m_c} + {15 \over 8} {\Delta m^2 \over m_c^2} + {5 \over
8} {\delta m^2 - q^2 \over m_c^2} \right ) . \end{eqnarray} 
The $B \to D_0 l \nu$ channel 
\begin{equation} {1 \over L(q^2)} {d\Gamma(B \to D_0 l\nu) \over dq^2} =
|\vec{q}|_{n=0} \left ( 1 - {\delta m \over m_c + m_d} + {3 \over 2} {\delta m^2
\over m_c^2} - {q^2 \over 2m_c^2} \right ) - {\rho_0^2 \over m_c^2}
|\vec{q}|_{n=0}^3 .
 \end{equation} 
using the definition~(\ref{rho}). 

The $B \to D_n l\nu$ ($n \not= 0$) channel: 
\begin{equation} {1 \over L(q^2)} {d\Gamma (B \to
D_n l \nu) \over dq^2} = {\rho_n^2 \over m_c^2} \left (
 (\delta m - \Delta_n)^2 - q^2
\right )^{3/2} \end{equation}
   Now everything is ready for the calculation of
the integrated SL decay rate. We again consider 
$L=(q^2)^N$. 

\subsubsection{The integrated rate and the global duality} 

It is convenient to represent the results for the partial decay rates in terms of
their ratios to the free quark decay rate. The latter has the form
\begin{eqnarray}
\Gamma(b\to c l \nu)&=&
(\delta m)^{2N+3}\left[ A^N_{1/2}\left( 1-\frac{3}{2}\frac{\delta m}{m_c}
+\frac{15}{8}\frac{\delta m^2}{m_c^2}\right)
+\frac{5}{8}A^N_{3/2}\frac{\delta m^2}{m_c^2}
+O\left(\frac{\delta m^3}{m_c^3}\right)\right].
\end{eqnarray}
Making use of the relation (\ref{a}) we find 
\begin{eqnarray} \frac{\Gamma(B\to D_0 l\nu)}{\Gamma(b\to c l\nu)}&=&
1-\frac{3\,\rho_0^2}{2N+5}\frac{\delta m^2}{m_c^2}
+\frac{3}{2}\,\frac{m_d}{1+m_d/m_c}\,\frac{\delta m}{m_c^2}
-(2N+3)\frac{\beta_0^2}{2m_c^2}\\ \frac{\Gamma(B\to D_n l\nu)}
{\Gamma(b\to c l\nu)}&=&
\frac{\delta m^2}{m_c^2}\,\frac{3\,\rho_n^2}{2N+5}\left(1-\frac{\Delta_n}{\delta
m}\right)^N\nonumber\\ &=&\frac{3\,\rho_n^2}{2N+5}\,\frac{\delta m^2}{m_c^2} -
3\left(\rho_n^2\Delta_n\right)\frac{\delta m}{m_c^2}
+\frac{1}{m_c^2}\sum_{k=2}^{2N+5}(-1)^k\frac{1}{2N+5}C^k_{2N+5}
\frac{\left(3\rho_n^2\Delta_n^k\right)}{\delta
m^{k-2}} 
\end{eqnarray}

Some remarks are in order: \par
\begin{itemize}
\item[1.] The main part of the OPE (i.e. the free quark decay) is reproduced by the
$\Gamma (B \to D_0 l\nu)$, within the leading and the subleading $1/m_c$ orders. The
excited states contribute only within the $(\delta m)^2/m_c^2$ and
$\Lambda \delta m/m_c^2$ orders in the SV limit. \par

\item[2.] Nevertheless, each of the individual exclusive channels contains
potentially large terms of the order $\delta m^2/m_c^2$ and 
$\Lambda\delta m/m_c^2$ which are absent in the OPE series.
\end{itemize} 

Now summing over all exclusive channels we find
\begin{eqnarray}\label{Xcsurc}
\frac{\Gamma(B\to X_c l\nu)}{\Gamma(b\to c l\nu)}&=&
1-\frac{\delta m^2}{m_c^2}\left(\rho_0^2-\sum_{n=1}^{n_{max}}\rho_n^2\right)\frac{3}{2N+5}
+3\frac{\delta m}{m_c^2}
\left(\frac{1}{2}\frac{m_d}{1+m_d/m_c}-\sum_{n=1}^{n_{max}}\rho_n^2\Delta_n\right)\nonumber\\
&&-(2N+3)\frac{\langle B|\vec k^{\,2}+V_1|B\rangle}{2m_c^2}
+(2N+4)\frac{\left( \sum_{n=1}^{n_{max}}\rho_n^2\Delta_n^2\right)}{2m_c^2}\nonumber\\
&&+\frac{1}{m_c^2\delta m}\sum_{k=3}^{2N+5}\frac{(-1)^kC^k_{2N+5}}{2N+5}
\frac{\left(3\sum_{n=1}^{n_{max}}\rho_n^2\Delta_n^k\right)}{\delta m^{k-3}}
\end{eqnarray}

The sum over the charm resonance levels is
truncated at $n_{max}$, which is the number of the resonance levels opened at 
$q^2=0$. For the confining potential and in the SV limit $n_{max}$ 
is found from the relation $\Delta_{n_{max}}\simeq \delta m$. 

Using the sum rules (\ref{sr0b})-(\ref{sr4b}) to rewrite the OPE result (\ref{finalope}) 
as the sum over hadronic resonances, the difference between the OPE and the
exclusive sum (the duality-violating contribution) explicitly reads 
\begin{eqnarray}
\label{dv}
\frac{\Gamma^{OPE}(B\to X_cl\nu)-\Gamma(B\to X_cl\nu)}{\Gamma(b\to cl\nu)}=
\frac{\delta m^2}{m_c^2}\sum_{k=0}^{2N+5}\frac{(-1)^{k}C^k_{2N+5}}{2N+5}
\frac{\delta^{(k)}}{\delta m^k}+O(\Lambda^2\delta m/m_c^3), 
\end{eqnarray}
where
\begin{equation}
\label{deltak}
\delta^{(k)} \equiv \sum_{n=n_{max}+1}^{\infty} \rho_n^2 (\Delta_n)^k=
\sum_{n=n_{max}+1}^{\infty}\left[r_n^2+O(\Lambda\delta m/m_c^2)\right](\Delta_n)^k. 
\end{equation}
As expected, this duality-violating contribution is connected with the charm 
resonance states forbidden kinematically in the decay process.
This kinematical truncation of the higher resonances induces a violation of duality 
equal to $\frac{(\delta m)^{2-k}}{m_c^2}\delta^{(k)}$ for every $k<2N+5$. 

To estimate the error induced by the truncation and thus the size of the 
duality-violation effects, we need to know the behavior of the
excitation energies and the transition radii at large $n$. 

\noindent 
1. 
For quite a general form of the confining potential we can write the 
following relations for $\Delta_n$ for large $n$ (recall that 
in the SV limit 
$\Delta_{n_{max}}\simeq \delta m$)  
\begin{eqnarray} 
\Delta_{n_{max}}=\Lambda C (n_{max})^a=\delta m, \nonumber \\
\Delta_{n}\ge\Lambda C n^a, \qquad n>n_{max},
\end{eqnarray}
with $C$ and $a$ some positive numbers. 
In particular, this estimate is valid for the confining potentials 
with a power behavior at large $r$. 

{\it This estimate for $\Delta_n$ is only depending on 
the behaviour of the potential at large distances (the infrared region)}. 

\vspace{.2cm}
\noindent 
2. 
The transition radii $r_n^2$ satisfy sum rules similar to sum rules for 
$\rho_n^2$ in Section V, namely\footnote{
Notice that these relations are exact and do not have any 
$1/m_c$ corrections.} 
\begin{eqnarray}
\label{srr}
&&\sum_{n=1}^{\infty} r_n^2=r_0^2,\nonumber\\
&&\sum_{n=1}^{\infty} r_n^2\Delta_n=\frac{m_d}{2(1+m_d/m_c)},\nonumber\\
&&\sum_{n=1}^{\infty} r_n^2(\Delta_n)^{(k+2)}
=\frac{1}{3}\langle D_0|k_j(h_{cd}-\epsilon_{D_0})^k k_j|D_0\rangle, 
\quad k=0,1,2,\dots. 
\end{eqnarray}
Hence, the behavior of the radii $r_n^2$ at large $n$ are connected with 
the finiteness of the r.h.s. of the sum rules.  
We can guarantee this for the Bjorken and Voloshin sum
rules, where finite values stand in the r.h.s. 
(the ground state radius $r_0$ is finite for the confining potential).  
In general, the finiteness of the matrix elements of the operators 
$k_j(h_{cd}-\epsilon_0)^k k_j$ 
(such as e.g. the kinetic energy of quarks in the ground state)  
depend on the properties of the potential at small $r$ 
({\it the ultraviolat behavior}) and probably also at large $r$ 
({\it the infrared behavior})
\footnote{We don't have a classical Wilsonian 
scheme where the ultraviolet region is referred to the Wilson coefficients 
and the infrared region is referred to the matrix elements of the operators, 
so we can have these regions mixed}. 
We have assumed throughout the paper that the average kinetic energy 
of the light spectator quark in the ground state is finite, i.e. 
$\langle D_0|\vec k^2|D_0\rangle\simeq \Lambda^2$. 
This already restricts some properties of the potential at small $r$ and 
provides convergency of one more sum rule in Eq. (\ref{srr}). 
If, in addition to this, we assume that the average values of the operators 
$k_j(h_{cd}-\epsilon_{D_0})^k k_j$ for $k=1,...,K$ over the ground state are 
finite, then combining with the behavior of the energies at large $n$ we come to
the following estimate 
\begin{eqnarray} 
r_n^2\lesssim \frac{1}{n^{1+\varepsilon}}\left(\frac{1}{n^a}\right)^{2+K}, 
\qquad \varepsilon >0. 
\end{eqnarray}
This allows us to obtain the duality-violation originating from the 
truncation of the various sum rules: 
\begin{eqnarray} 
\label{accuracyBV} 
{\rm Bjorken}:  \qquad \frac{\delta m^2}{m_c^2}\delta^{(0)}
&=&\frac{\delta m^2}{m_c^2}\sum_{n_{max}} r_n^2
\lesssim \frac{\delta m^2}{m_c^2}\left(\frac{1}{n_{max}^{a}}\right)^{K+2}
\simeq\frac{\Lambda^2}{m_c^2}\left(\frac{\Lambda}{\delta m}\right)^{K}
\nonumber\\
{\rm Voloshin}: \qquad \frac{\delta m^2}{m_c^2}\delta^{(1)}
&=&\frac{\delta m}{m_c^2}\sum_{n_{max}} r_n^2 \Delta_n
\lesssim \frac{\delta m}{m_c^2}\Lambda 
\left(\frac{1}{n_{max}^a}\right)^{K+1}
\simeq\frac{\Lambda^2}{m_c^2}\left(\frac{\Lambda}{\delta m}\right)^{K}.
\end{eqnarray}
Similar estimates can be done for higher moment sum rules.  
One can see that the truncation error in any of the sum rules leads to the 
duality-violation of the same order 
$O\left(\Lambda^{2+K}/m_c^2\delta m^{K}\right)$. 
An interesting feature about these estimates is that the dependence on $a$ 
has disappeared from the final result. 
Hence, {\it the estimates are independent 
of the details of the potential at large} $r$, provided the potential 
guarantees the confinement, i.e. $a$ is positive.  

These are however rather crude estimates which do not take into account further 
possible suppressions (due e.g. to the orthogonality of the wave functions of the 
ground $n=0$ and the excited states $n > 0$). In such a case  
the real accuracy is better, and might depend on the 
details of the potential also at large $r$. In general, we can state that 
the truncation (duality-violaiton) error occurs at the order 
\begin{eqnarray} 
\frac{\Lambda^2}{m_c^2}\left(\frac{\Lambda}{\delta m}\right)^b,  
\end{eqnarray}
where the exponent $b>0$ depends on the properties of the potential (in general, 
both at short and long distances). 
A more detailed analysis of which potentials satisfy the above requirements is 
beyond the scope of this paper and is left for another publication \cite{15r}. 

If we would like to have the truncation error of a higher order in $1/m_c$, 
e.g. in $O(\Lambda^3/m_c^3)$, this is not so straight. Namely, in this case we need 
\begin{eqnarray}
\label{mc3}
\frac{\delta^{(k)}}{\delta m^k}
\simeq O\left(\frac{\Lambda^3}{\delta m^2m_c}\right). 
\end{eqnarray}  
As we have noticed, the series for $\delta^{(k)}$ in the main part does not 
depend on $m_c$, so the only possibility to have the relation (\ref{mc3}) fulfilled 
in the SV limit, is to have $r_n^2=0$ starting from some number $n$. 
(Exactly this situation takes place for the HO potential where all 
$D_0\to D_n$ transition radii for $n > 1$ 
are equal zero \cite{16r}). In this case for large enough values of
$\delta m$, the term proportional to $r_n^2$ in (\ref{deltak}) disappears and the 
second term provides the truncation error of the order 
$O(\delta m^2\Lambda^2/m_c^4)$. 

As we are going to show elsewhere, the accuracy of duality of order $\Lambda^3/m_c^3$ 
can be achieved if we keep a fixed ratio $\delta m/m_c$ when $m_c\to\infty$. 
One can proceed exactly along the same lines, but technically a bit different 
treatment is necessary: namely, at several places throughout the paper terms of the order 
$\delta m^3/m_c^3$ have been omitted, and they should be kept if the limit 
$\delta m/m_c=const$ is considered. This analysis will be presented in \cite{15r}. 

Finally, it is interesting to notice that all resonance levels opened at $ q^{2} =
0$ are contributing on equal footing to the sum rules and therefore to the decay
rate. So, a considerable delay in opening channels with large $n$ compared to
the channels with small $n$ with the increasing recoil does not matter at all. 
This is a very important feature which basically determines a high accuracy of 
the OPE calculation of the integrated decay rate (cf. \cite{isgur,nous}).


\subsubsection{The smeared $q^2$-distribution and the local duality} 
The situation however differs considerably if we consider the differential decay
widths. We find it more physical to use here the four vector $q^2$ variable.
The region near $q_{max}^2$ (zero recoil) is special: as we move to higher $q^2$, 
the excited channels close one after another leaving ultimately only
the $D_0$ ground state opened. 

Let us consider a partially integrated decay rate in the $q^2$-region 
above the threshold of the $D_{n=1}$ channel. 
In this case the relation between the OPE and the exact result 
(which is reduced in this case to the exclusive $B\to D_0 l\nu$ decay) reads: 
\begin{eqnarray}
\label{local}
\int\limits^{\delta m^2}_{(\delta m-\Delta_1)^2}
dq^2\frac{d\Gamma(B\to X_cl\nu)}{dq^2}=
\int\limits^{\delta m^2}_{(\delta m-\Delta_1)^2}
dq^2\frac{d\Gamma(b\to cl\nu)}{dq^2}\left[1+O\left(\frac{\Delta_1\delta m}{m_c^2}\right)\right]. 
\end{eqnarray}
In this formula we have neglected a difference between the upper boundaries of
the quark and  hadron channels of the order $\Lambda^2/m_c^2$.   Eq.
(\ref{local}) means that local duality near maximal $q^2$ is
violated at order $O(\Lambda_1\delta m/m_c^2)$. As we have seen, 
the dangerous terms of
this order are cancelled in the integrated rate  against similar contributions of
other channels due to the Voloshin sum rule. However, the 
$O(\Lambda\delta m/m_c^2)$
violation of the local duality might have negative
consequences for the application of the method to the analysis of the
experimental results. For example this happens if one observes only a small part of the 
phase space near maximal $q^2$ \cite{isgur}. 

\section{Conclusion}
We have studied quark-hadron duality in decays of heavy mesons 
in the SV limit using the nonrelativistic potential model for the description 
of mesons as $q\bar q$ bound states. Our main results are as follows:
\begin{itemize}
\item[1.] 
The OPE is constructed and the following $\Lambda/m_c$ and $\Lambda/\delta m$
double series is found for the integrated decay rate:     
\begin {eqnarray}
\label{mainresult}
\frac{\Gamma^{OPE}(B\to X_cl\nu)}{\Gamma(b\to cl\nu)}
=1+C_0\frac{\langle B|\vec k^{\,2}+V_1|B \rangle}{2m_c^2}
 +(1-C_0)\frac{\langle B|\vec k^{\,2}|B \rangle}{2m_c^2}
 +\sum_{k=1}^{k_{0}}C_k\frac{\langle B|\vec k(h_{bd}-\epsilon_B)^k \vec k|B \rangle}
 {2m_c^2(\delta m)^k}
 +O\left(\frac{\Lambda^2\delta m}{m_c^3}\right),\nonumber 
\end{eqnarray}
where $C_k$ are calculable constants and $k_{0}$ depends on the leptonic tensor. 

\item[2.] The HQ expansion of the transition form factors in the nonrelativistic
potential model is performed. A nonrelativistic analog of the Luke theorem for the
exclusive transition form factor between the ground states is obtained.

It is shown that the sum of the squares of 
the $B \to D_n l\nu$ transition form factors are expressed through the 
expectation values of the operators emerging in the OPE series. These
nonrelativistic analogs of the Bjorken, Voloshin, and higher order sum rules
provide a bridge between the sum over exclusive channels and the OPE series. 

\item[3.] The integrated decay rate is calculated by direct summation of
exclusive channels. For the comparison of this directly calculated 
$\Gamma (B \to X_cl\nu)$ and the corresponding $\Gamma^{OPE}(B \to X_cl\nu)$ 
the sum rules are necessary. A difference (duality-violation) between 
the two expressions is observed. As shown explicitly by the use of the sum 
rules, this difference is connected with the higher $c\bar{d}$ resonances 
which are forbidden kinematically in the decay process but are implicitly 
taken into account in the OPE approach. 
Therefore the accuracy of the OPE is directly related to the error induced 
by the kinematical truncation in the sum rules (Bjorken, Voloshin, etc). 
The actual error depends on the convergence of the series, i.e. on the nature of the
potential. We have discussed the constraints on the latter convergence 
which lead to the duality violation of order $O(\Lambda^{2+b}/m_c^2\delta m^b)$ with 
$b$ depending on the behavior of the potential both at the short and long 
distances. 

{\it Up to the mentioned duality-violation}, the agreement between the 
OPE and the exclusive sum is achieved within different $1/m_c$ orders 
due to different reasons:

The leading order and the subleading $\delta m/m_c$ and
$\Lambda/m_c$ orders the free quark integrated decay
rate $\Gamma (b \to c l\nu)$ is equal to the rate of the transition into 
the ground state $D_0$. This is due to the specific behavior of the transition form factor between the
ground states near the zero recoil (Luke theorem). Also part of the 
$\Lambda^2/m_Q^2$ correction in the OPE result proportional to the 
$\langle B|\vec{k}^2 + V_1|B \rangle$ matches to the
contribution of the ground state $D_0$ in the exclusive sum.

For higher order terms the agreement between the OPE and the exclusive sum is a
collective effect due to subtle cancellations in the sum over exclusive channels:

Namely, each of the individual decay rates $\Gamma (B \to D_nl\nu)$ contain
potentially large terms of the order $\delta m^2/m_c^2$ and $\Lambda\delta m/m_c^2$.
These terms cancel in the exclusive sum due to the Bjorken and Voloshin sum
rules, respectively. The higher order sum rules allow us to represent the
contribution of exclusive channels in terms of the average values of the
operators $O_i$ over the $B$-meson state.

\item[4.] 
If the differential semileptonic decay widths are considered near maximum $q^2$, 
the violation  of the local duality occurs at order $O(\Lambda\delta m/m_c^2)$.
\end{itemize}

Clearly, in QCD the situation is more complicated because of the 
multiparticle $X_c$ states, pion emission, hybrid and multiquark exotic $D$ mesons, 
radiative corrections. Nevertheless the 
duality violation due to the kinematical truncation of the series should be 
quite similar to the case of nonrelativistic quantum mechanics.
Also similar is the role of the inclusive sum rules in obtaining the duality 
relations. 

\acknowledgements 
The authors are grateful to G. Korchemsky, S. Simula, B. Stech and A. Vainshtein 
for fruitful discussions. We are particularly indebted to B. Stech for a careful 
reading of the paper before publication. A.L., L.O., O.P. and J.-C. R. would like 
to acknowledge partial support from the EEC-TMR Program, contract No. CT 98-0169.


\begin{thebibliography}{30}
\bibitem{cgg} J. Chay, H. Georgi, B. Grinstein,  
Phys. Lett. B {\bf 247}, 399 (1990). 
\bibitem{bsuv93} I. Bigi, M. Shifman, N. Uraltsev, A. Vainshtein, 
Phys. Rev. Lett. {\bf 71}, 496 (1993), Phys. Rev. D {\bf 49}, 3356 (1994). 
\bibitem{mw} A. Manohar and M. Wise, Phys. Rev. D {\bf 49}, 1310 (1994). 
\bibitem{fls} A. Falk, M. Luke, and M. Savage, Phys. Rev. D {\bf 49}, 3367 (1994). 
\bibitem{shifman} A comprehensive discussion of duality in heavy meson
decays can be found in 
B. Chibisov, R. Dikeman, M. Shifman, N. Uraltsev, Int. J. Mod. Phys. A {\bf 12}, 
2075 (1997); 
B. Blok, M. Shifman, D.-X. Zhang, Phys. Rev. D {\bf 57}, 2691 (1998).
\bibitem{isgur} N. Isgur, Phys. Lett. B {\bf 448}, 111 (1999).
\bibitem{sv} M. B. Voloshin and M. A. Shifman, Yad. Fiz. {\bf 47}, 801 (1988)
[Sov. J. Nucl. Phys. {\bf 47}, 511 (1988)]. 
\bibitem{thooft} G. 't Hooft, Nucl. Phys. {\bf B75}, 461 (1974).
\bibitem{gl} B. Grinstein and R. Lebed, 
Phys. Rev. D {\bf 57}, 1366 (1998).
\bibitem{b-thooft} I. Bigi, M. Shifman, N. Uraltsev, A. Vainshtein, 
Phys. Rev. D {\bf 59}, 054011 (1999).
\bibitem{bgm} G. Boyd, B. Grinstein, and A. Manohar, 
Phys. Rev. D {\bf 54}, 2081 (1996).
\bibitem{nous}A. Le Yaouanc, D. Melikhov, V. Mor\'enas, L. Oliver, O. P\`ene and
J.-C. Raynal, Phys. Lett. B {\bf 480}, 119 (2000).  
\bibitem{15r} A. Le Yaouanc {\it et al.}, paper in preparation.
\bibitem{13r} J. D. Bjorken, invited talk at Les Rencontres de Physique de la
Vall\'ee d'Aoste, La Thuile, Italy, SLAC Report No. SLAC-PUB-5278 (1990)
unpublished~; N. Isgur and M. Wise, Phys. Rev. D {\bf 43}, 819  (1991). 
\bibitem{14r} M. B. Voloshin, Phys. Rev. D {\bf 46}, 3062 (1992).
\bibitem{greg} A.G. Grozin and G.P. Korchemsky, Phys. Rev. D {\bf 53} (1996) 1378.
\bibitem{16r} A. Le Yaouanc {\it et al.}, preprint 
hep-ph/0005039. 
\bibitem{luke} M. Luke, Phys. Lett. B {\bf 252}, 447 (1990).
\end{thebibliography}
\end{document}